\documentstyle[psfig]{mn}

\psfigurepath{/z/rue/dima/dima}

\def\bl{\par\vskip 12pt\noindent}
\def\Pm{\mathop{\rm Pm}\nolimits}
\def\q{\qquad}
\def\bib{\item}
\def\beg{\begin{eqnarray}}
\def\ende{\end{eqnarray}}
\def\ass{$\alpha_{\rm SS}$}

\title[Jet launching  and the vertical structure of  accretion discs]
      {Jet launching theory and the vertical structure of a magnetically-dominated thin accretion disc}
\author[G.~R\"udiger and D. Shalybkov]
       {G.~R\"udiger$^1$ and D. Shalybkov$^{1,2}$
       \\ $^1$ Astrophysikalisches Institut Potsdam,
         An der Sternwarte 16, D-14482 Potsdam, Germany
       \\ \q gruediger@aip.de
       \\ $^2$ A.F. Ioffe Institute for Physics and Technology,
         194021, St. Petersburg, Russia
       \\ \q dasha@astro.ioffe.rssi.ru }

\date{\today}

\begin{document}

\maketitle

\begin{abstract}
The presence of an imposed external magnetic  field may drastically 
influence the structure of thin accretion discs. The magnetic field
energy is here assumed to be in balance with the thermal energy of
the accretion flow. 
The vertical magnetic field, its toroidal component $B^{\rm tor}$ at the disc
surface (due to different  rotation rates between disc and its magnetosphere),  
the turbulent magnetic Prandtl number and the  viscosity-alpha
are the key parameters of our model.  
Inside the corotation radius for rather small $B^{\rm tor}$ the resulting inclination angle $i$ of the
magnetic field lines to the disc surface normal can exceed the critical value
30$^\circ$ (required to launch  cold jets) even for small magnetic Prandtl numbers
of order unity. The self-consistent consideration of both magnetic field and
accretion flow demonstrates a weak dependence of the inclination (``dragging'') 
angle on the 
magnetic Prandtl number for given surface density but a strong dependence
on the toroidal field component at the disc surface. 
 
A magnetic disc is thicker than a nonmagnetic one for typical parameter
values. The accretion rate can be strongly amplified by large $B^{\rm tor}$ 
and small magnetic Prandtl number. On the other hand, for given accretion rate 
the magnetised disc is less massive than the standard-alpha disc.  
The surface values of the toroidal magnetic fields which are necessary to
induce considerably high values for the inclination angle are  much
smaller than expected and are of order $10^{-3}$ of the imposed vertical field.
As the innermost part of the disc produces the largest $B^{\rm tor}$, the 
largest radial inclination can be expected also there. The idea is therefore 
supported that the cold jets are launched only in the central disc area.
\end{abstract}
\section{Introduction} 
Accretion flows is assumed to occur in a variety of astrophysical
objects. The structure of magnetic accretion discs is the problem 
of major importance, for example, for understanding of the origin of
astrophysical jets. If the inclination, $i$,   of the field lines from the
vertical exceeds 30$^\circ$, the plasma can be accelerated when spiralling
along the field line (Blandford \& Payne 1982; Lynden-Bell 1996; Campbell 
1997; Krasnopolsky, Li \& Blandford 1999).
Accretion disc-driven magnetocentrifugal winds have been widely used
to model the astrophysical jets (e.g. Ustyugova et al. 1999, Ouyed
\& Pudritz 1999, Krasnopolsky, Li \& Blandford 1999).

The present study is motivated by a series of papers dealing
with the interaction of accretion discs with external magnetic
fields (Livio \& Pringle 1992; Lubow, Papaloizou \& Pringle 1994;
Bardou \& Heyvaerts 1996; Reyes-Ruiz \& Stepinski 1996; Ogilvie 1997;
Ogilvie \& Livio 1998; Campbell 1998; Campbell \& Heptinstall 1998a, 1998b;
Brandenburg \& Campbell 1998). In the presence of a vertical magnetic field,
the vertical gradient of angular velocity generates the toroidal field due 
to the stretching effect. This field may generally influence the vertical 
and radial structure of a disc and alter the angular momentum transport.
If the magnetic field is sufficiently strong it even will influence the
rotation law (Ogilvie 1997; Ogilvie \& Livio 1998). 

Our previous paper (Shalybkov \& R\"udiger 2000, hereafter paper I))
considered the self-consistent steady-state 
structure of the polytropic accretion disc in the presence of a vertical
magnetic field. In the present paper we will refuse from simplified
polytropic equation of state and solve the full system of
MHD equations for the vertical accretion disc structure. So better comparisons
with the theory of the vertical structure of accretion discs without magnetic 
field are possible. As known, such computations, 
for given viscosity-alpha and given opacity law,
yield the accretion rate $\dot M$ for any possible column density $\Sigma$. 
The same is done here for an accretion disc threaded by a magnetic field with a
given vertical component $B^{\rm vert}$ and a given toroidal component
$B^{\rm tor}$ which, of course, have opposite signs 
if we are within the corotation radius\footnote{At the corotation radius the 
    component $B^{\rm tor}$ vanishes; 
for a nonrotating central object the corotation radius is in the infinity}.
Such a configuration can only exist for a certain accretion rate and with a 
certain distribution of a radial magnetic field component $B_R$.
The value of the latter taken at the surface defines the inclination
angle $i$ known from the jet theory. Along this way the well-known 
dragging problem -- to find the {\rm radial} inclination $i$ -- has been
unified  with the theory of the vertical structure of accretion discs.  

There is an estimate for  the toroidal component $B^{\rm tor}$ due to
the different rotation of halo and disc.
If a disc  halo with a high conductivity exists then the surface
value of $B_\phi$ will become large. After
Campbell (1992) the shear between the rigidly rotating halo and
the accretion disc induces a toroidal magnetic field of 
\beg
B^{\rm tor} = - \gamma {R \over H} {{\rm Pm} \over \alpha_{\rm SS}}
\ {\Omega_{\rm Kep} - \Omega_* \over \Omega_{\rm Kep}}  B^{\rm vert}
\label{camp}
\ende
with $\Omega_*$ as the stellar rotation rate.
The numerical integration of the induction equation 
for a uniform  and a dipolar field and a plasma halo (with conductivity of 10 
times of the disc conductivity) confirms this result with 
$\gamma \simeq 1$ (Elstner \& R\"udiger 2000). 
The toroidal surface field changes 
its sign at the corotation radius where $\Omega_{\rm Kep}$ equals
the stellar rotation rate $\Omega_*$. 
The magnetic torque results as negative inside the corotation
radius and positive outside the corotation radius.
For a disc embedded in vacuum, of course,  the $\gamma$ in (\ref{camp}) 
vanishes. We consider the $\gamma$ as representing the unknown halo conductivity.
We shall show  here that we only need very small $\gamma$ in order to produce 
inclination angels $i$ of the interesting value of 30$^\circ$ and
more -- independent of the the magnetic Prandtl number Pm. We also can take 
from (\ref{camp}) and  from the simulations by Elstner \& R\"udiger that the 
ratio
\beg
\beta= {B^{\rm tor}\over B^{\rm vert}}
\label{beta}
\ende
-- which is negative (positive) inside (outside) the corotation radius grows (by absolute value) inwards. This fact will play an important role in
the philosophy of the presented research.
\section{Basic equations} 

Consider the structure of an axisymmetric  disc in a steady-state regime; 
$R$, $\phi$, and $z$ are the cylindrical coordinates. The resulting kinetic
and magnetic equations are given in paper I as  the Eqs. (1)...(13) there.
Here we have only to add the energy equation.
All generated energy is assumed to be transfered from the disc by radiation
neglecting convection, and the disc is optically thick. Then the energy
equation is
\beg
\lefteqn{{1 \over R} {\partial \over \partial R}(R F_R) +
{\partial F_z \over \partial z} =
\rho\nu \left[ 2\left( \frac{\partial u_R}{\partial R} \right)^2
+ 2\left( {u_R \over R} \right)^2 + \right. } \nonumber \\
& & \left. + 2\left( {\partial u_z \over
\partial z} \right)^2 +  \left( R { \partial \over \partial R}
\left( {u_\phi \over R} \right)
\right)^2 + \right. \nonumber \\
& & \left. + \left( {\partial u_\phi \over \partial z} \right)^2
+ \left( {\partial u_R \over \partial z} + {\partial u_z \over
\partial R} \right)^2 -  \right.  \nonumber \\
& & \left. - {2 \over 3} \left( \frac{1}{R} 
\frac{\partial}{\partial R} (R u_R) + \frac{\partial u_z}{\partial z} \right)^2
\right] + {\eta_{\rm T} \over \mu_0} \left[ 
\left( {\partial B_\phi \over \partial z} \right)^2 + \right. \nonumber \\
& & \left. + \left(
{\partial B_R \over \partial z} - {\partial B_z \over \partial r}
\right)^2 + \left( {1 \over R} { \partial \over \partial R}(RB_\phi)
\right)^2 \right],
\ende
where the energy flux components are
\beg
F_R=-{16 \sigma T^3 \over 3 \kappa \rho}
{\partial T \over \partial R},
\q
F_z=-{16 \sigma T^3 \over 3 \kappa \rho}
{\partial T \over \partial z}
\ende
 with
$\sigma$ is the Stefan-Boltzmann constant, $T$ is the temperature
and $\kappa$  the Roseland mean opacity. 

The equations  must be supplemented by
relations specifying the gravitational potential, the equation of state,
the opacity, the viscosity, and the magnetic diffusivity. 
We neglect the self-gravitation of the disc so that 
\beg
\psi=-{{\rm G} M_* \over (R^2+z^2)^{1/2} },
\ende
where G is the gravity constant and  $M_*$ is the mass of the central object. 
We  adopt here the ideal-gas equation of state 
$
P={\cal R}\rho T/\mu 
$
(${\cal R}$ is the molar gas constant and $\mu$ is the mean molecular
mass) and  the opacity may fulfil a power law, 
$
\kappa=k_0 \rho^\gamma T^\delta,
$
with the constant quantities  $k_0$, $\gamma$ and $\delta$. This includes
the cases of Thomson scattering opacity ($k_0 \approx 0.4$, $\gamma=\delta=0$)
and Kramers opacity ($k_0 \approx 6.6 \cdot 10^{22}$, $\gamma=1$,
$\delta=-3.5$). A Shakura-Sunyaev parameterisation is used for the
turbulent viscosity,
\beg
\rho \nu_{\rm T}= \alpha_{\rm SS} {P \over \Omega},
\ende
where $\alpha_{\rm SS}$ is a constant (Shakura \& Sunyaev 1973).
For the magnetic diffusivity we assume that the magnetic Prandtl
number
$
\Pm = {\nu_{\rm T} / \eta_{\rm T}} 
$
is constant.

Following Regev (1983) and Kluzniak \& Kita (2000) we scale all
quantities by their correspondent characteristic values. This
will make the equations dimensionless and allows to compare  the
relative significances of each term. The radial
distances are scaled by some characteristic radius, $\tilde R$,
and vertical distances by a typical vertical height of the disc,
$\tilde H$. We represent the angular velocity in units of the Keplerian
velocity at the characteristic radius, $\tilde \Omega^2={\rm G}M_*/\tilde{R}^3$.
Using these three characteristic quantities all
others can be defined. The typical sound speed is $\tilde{c}_{\rm s}=\tilde H \tilde \Omega$,
the typical viscosity and magnetic diffusivity $\tilde \nu=\tilde \eta=
\tilde{c}_{\rm s} \tilde H$, the typical temperature is $\tilde{T}={ \mu /
{\cal R} \tilde{c}_{\rm s}^2}$, the typical pressure is $\tilde{P}=\tilde{\rho}
\tilde{c}^2_{\rm s}$,
the typical energy flux is $\tilde{F}=\tilde{\rho}\tilde{c}_{\rm s}^3$, and
the typical magnetic field is
$\tilde{B}=(\mu_{0}\tilde{\rho})^{1/2}\tilde{c}_{\rm s}$.
The last three quantities use typical density, $\tilde \rho$,
which transforms the equation for the vertical energy flux
to the dimensionless form, i.e.
\beg
\tilde{\rho}= \left( { 16 \sigma \over 3 k_0 } \tilde{H}^{4-2\delta}
\tilde{\Omega}^{5-2\delta} \left( {\mu \over {\cal R}} \right)^{4-\delta}
\right)^{1 \over \gamma + 2}. 
\label{dens}
\ende
Note that the typical magnetic field defined above $\tilde{B}$ gives
$\tilde V_A=\tilde{B}/(\mu_{0}\tilde{\rho})^{1/2} \equiv \tilde{c}_{\rm s}$,
where $\tilde V_A$ is the typical Alfv\'{e}n velocity. The last relation 
defines the magnetic field energy as  in  balance with the accretion
flow energy. The dimensionless equation of state takes the simple form
$
P=\rho T, 
$
where the same symbols as before are used for the dimensionless quantities. 

Let us define the parameter $\epsilon$  
\beg
\epsilon = {\tilde H / \tilde R}={V_A / \tilde \Omega  \tilde R}.
\label{eps} 
\ende
As in paper I we consider a geometrically thin disc ($\epsilon \ll 1$)
and expand all variables in power of $\epsilon$, i.e.
\begin{eqnarray*}
\lefteqn{A(R,z)=A_0(R,z)+\epsilon A_1(R,z)+....}
\end{eqnarray*}
Using the same symbols for the normalized quantities,  
the full system at the leading order of $\epsilon$ takes the form
\beg 
u_{R0} B^{\rm vert} + \eta_{\rm T0} \frac{\partial B_{R0}}{\partial z} =0,
\label{e1}
\ende
\beg 
R B^{\rm vert}\frac{\partial \Omega_1}{\partial z}+ R B_{R0} 
\frac{\partial \Omega_0}{\partial R} + \frac{\partial}{\partial z}
\left( \eta_{\rm T0} \frac{\partial B_{\phi 0}}{\partial z} 
\right)=0, 
\label{e2}
\ende 
\beg
-2 \rho_0 \Omega_0 \Omega_1 R = 
B^{\rm vert} \frac{\partial B_{R0}}{\partial z} 
+\frac{\partial}{\partial z} \left(
\rho_0\nu_{\rm T0} \frac{\partial u_{R0}}{\partial z} 
\right),
\label{e3}
\ende 
\beg 
\rho_0 {u_{R0} \over R} {\partial \over
\partial R}(R^2 \Omega_0) =  
B^{\rm vert} \frac{\partial B_{\phi0}}{\partial z} + 
\frac{\partial}{\partial z} \left(
\rho_0\nu_{\rm T0} R {\partial \Omega_1 \over \partial z} \right),
\label{e4}
\ende 
\beg 
\frac{\partial P_0}{\partial z}+\rho_0 {z \over R^3}
+ B_{\phi 0}\frac{\partial B_{\phi 0}}{\partial z} 
+B_{R0} \frac{\partial B_{R0}}{\partial z}=0,
\label{e5}
\ende
\beg
\lefteqn{{\partial F_0 \over \partial z}=
\rho_0\nu_{\rm T0} \left[ 
\left( R { \partial \Omega_0 \over \partial R} \right)^2
 + \left( R{\partial \Omega_1 \over \partial z} \right)^2
+ \left( {\partial u_{R0} \over \partial z} \right)^2  \right]+ }\nonumber \\
& & + \eta_{\rm T0} \left[ \left( {\partial B_{\phi 0} \over \partial z} 
\right)^2 + \left( {\partial B_{R0} \over \partial z} \right)^2 \right]
\label{e6}
\ende
and
\beg
 F_0= - 
{ T_0^{3-\delta} \over \rho_0^{1+\gamma} }
{\partial T \over \partial z},
\label{e7}
\ende
where $B_{z0}\equiv B^{\rm vert}$ and $\Omega_0 = R^{-3/2}$ is the
normalized  Keplerian velocity. Equations (\ref{e1})....(\ref{e5}) are
identic with those of paper I. 
Note that after (\ref{e6}) and (\ref{e7}) all energy is assumed to be transported in
the vertical direction.

The system (\ref{e1})....(\ref{e7}) depends on $R$ only as parameter and we
can use the real radius $R$ as scaling constant $\tilde R$, the real
disc half-thickness $H$ as scaling constant $\tilde H$ and the real
Keplerian angular velocity $\Omega_{\rm Kep}=\sqrt{GM_*/R^3}$ as scaling
constant $\tilde \Omega$. The  system takes thus the final form
\beg 
u_{R} B^{\rm vert} + {\alpha_{\rm SS} \over \Pm}
T\frac{\partial B_{R}}{\partial z} =0, 
\label{uR0}
\ende
\beg 
B^{\rm vert}\frac{\partial u_{\phi}}{\partial z} - {3 \over 2} B_{R} 
+ {\alpha_{\rm SS} \over \Pm} \frac{\partial}{\partial z}
\left( T \frac{\partial B_{\phi}}{\partial z} 
\right)=0,
\ende 
\beg
-2 {P \over T} u_{\phi} = 
B^{\rm vert} \frac{\partial B_{R}}{\partial z} 
+ \alpha_{\rm SS} \frac{\partial}{\partial z} \left(
 P \frac{\partial u_{R}}{\partial z} 
\right),
\end{eqnarray} 
\beg 
0.5 {P \over T} u_{R} =  
B^{\rm vert} \frac{\partial B_{\phi}}{\partial z} + 
\alpha_{\rm SS} \frac{\partial}{\partial z} \left(
P{\partial u_{\phi} \over \partial z} \right),
\label{phi}
\ende 
\beg 
\frac{\partial P}{\partial z}+ {P \over T} z 
+ B_{\phi}\frac{\partial B_{\phi}}{\partial z} 
+B_{R} \frac{\partial B_{R}}{\partial z}=0, 
\ende
\beg
\lefteqn{{\partial F \over \partial z}=
\alpha_{\rm SS} P \left[ 
 { 9 \over 4}
 + \left( {\partial u_{\phi} \over \partial z} \right)^2
+ \left( {\partial u_{R} \over \partial z} \right)^2  \right]+ }\nonumber \\
& & + {\alpha_{\rm SS} \over \Pm} T
\left[ \left( {\partial B_{\phi} \over \partial z} \right)^2 
+ \left( {\partial B_{R} \over \partial z} \right)^2 \right]
\ende
and
\beg
F= - { T^{4-\delta+\gamma} \over P^{1+\gamma} }
{\partial T \over \partial z} ,
\label{F0}
\ende
if also the suffices 0 and 1 are dropped.
These relations  form a nonlinear set of differential equations
for $P$, $T$, $F$, $u_{R}$, $u_{\phi}$, $B_{R}$, $B_{\phi}$ with
$\alpha_{\rm SS}$, Pm, and $B^{\rm vert}$  as parameters. There are also two
implicit  parameters, i.e.  $R$ and the central mass $M_*$.
Solving the equations for given $R$ and $M_*$ the vertical 
structure for given $\alpha_{\rm SS}$, Pm and $B^{\rm vert}$ is found with 
10 reasonable  boundary conditions.

For discs symmetric with respect to the midplane ($z=0$) the physical
quantities such as $\Omega$, $u_R$, $T$ are even functions of $z$
while $B_R$, $B_{\phi}$, $F$ are odd functions of $z$. Such
symmetries provide the boundary conditions at the midplane ($z=0$) as
\beg
B_{R}=B_{\phi}={\partial u_{R}/\partial z} = 
{\partial u_{\phi} / \partial z}= F=0.
\label{b0} 
\ende 
At the disc surface, $z=1$, it no external pressure is allowed. The simplest boundary condition for the temperature, $T(1)=0$, 
is  used.  It reflects the fact that for
optically thick disc the surface temperature must be much smaller than
the temperature at the disc midplane. Then the first of the above equations 
immediately gives $u_{R}(1)=0$ for $T(1)=0$.
We additionally fix the toroidal and vertical magnetic field at the surface
as the last two boundary conditions. It means that the angle $\beta$ is fixed 
 and also the accretion rate (see (\ref{mnd})
and (\ref{cond})). Hence, the boundary conditions at $z=1$ are
\beg
P=T=u_{R}=0, \;
B_{\phi}=B^{\rm tor}, \; B_z=B^{\rm vert},
\label{b1}
\ende
where $B^{\rm tor}$ and $B^{\rm vert}$ are free constants.
As usual in the theory of the vertical structure of accretion discs, 
the accretion rate (or the column density $\Sigma$) remain the only free 
parameter.
\section{Physical parameters} 
It is useful to connect our normalized quantities with the real physical
values. The real physical parameters of the problem are the distance to
the central object, $R$, its mass $M_*$, the accretion rate $\dot M$ and
the inclination angle $i$ of the magnetic field lines to the rotation axis.
Note that we can choose any of the accretion rate, the surface density and
the disc half-thickness as the free parameter of the problem. All these quantities
are connected with each other (see below). Accretion rate is
more convenient as a parameter from the physical point of view, but the disc
half-thickness is more natural for numerical calculations.

The accretion rate $\dot M$ and the surface density $\Sigma$ as
functions of half-thickness $H$ are
\beg
\dot M=4\pi R \tilde \rho \tilde{c}_{\rm s} H \hat {\dot M},
\label{Mdot}
\ende
where the $\tilde \rho$ is given by (\ref{dens}) and the normalized  
accretion rate $\hat {\dot M}$ is
\beg
\hat{\dot M} =- \int_0^1 u_{R0} \rho_0 dz 
\label{mnd}
\ende
and
\beg
\Sigma=2\tilde \rho H \hat \Sigma ,
\label{sigma}
\ende
where the normalized surface density is
\beg
\hat \Sigma=\int_0^1 \rho_0 dz.
\ende
The disc thickness and the surface density as functions of
$\dot M$ are
\beg
H= \left( {3k_0 \over 16 \sigma} \left( {{\cal R} \over \mu} 
\right)^{4-\delta} \Omega_{\rm Kep}^{2\delta-7-\gamma} 
\left( {\dot M \epsilon \over
4 \pi \hat {\dot M}} \right)^{\gamma+2} \right)^{1 \over 10-2\delta+3\gamma} ,
\label{Hm}
\ende
where the $\epsilon$ after (\ref{eps}) is 
\beg
\lefteqn{\epsilon=}\nonumber\\
\lefteqn{\left( R^{2\delta-10-3\gamma} {3k_0 \over 16 \sigma}
\left( {{\cal R} \over \mu} 
\right)^{4-\delta} \Omega_{\rm Kep}^{2\delta-7-\gamma}
 \left( {\dot M \over
4 \pi \hat {\dot M}} \right)^{\gamma+2} \right)^
{1 \over 8-2\delta+2\gamma}}
\label{em}
\ende
and the surface density
\beg
\lefteqn{\Sigma=}\nonumber\\
\lefteqn{ 2\hat \Sigma \left( \left( {16\sigma \over 3k_0}
\left( {\mu \over {\cal R} } \right)^{4-\delta} \right)^2 
\Omega_{\rm Kep}^{4-2\delta-\gamma} \left( {\dot M \epsilon \over 4\pi 
\hat{\dot M}}
\right)^{6-2\delta+3\gamma} \right)^{1 \over 10-2\delta+3\gamma} . }
\ende
Half-thickness and accretion rate as functions of $\Sigma$ are
\beg
H=\left( {3k_0 \over 16 \sigma} \left( { {\cal R} \over \mu}
\right)^{4-\delta} \Omega_{\rm Kep}^{2\delta -5} \left(
{\Sigma \over 2\hat \Sigma} \right)^{2+\gamma}
\right)^{1 \over 6+\gamma-2\delta}
\ende
and
\beg
\dot M=4\pi R^2 \epsilon \Omega_{\rm Kep} \Sigma {\hat {\dot M} \over
2\hat \Sigma}
\ende
with  
\beg
\epsilon={1 \over R} \left( {3k_0 \over 16\sigma }
\left( {\Sigma \over 2 \hat \Sigma} \right)^{2+\gamma}
\left( { {\cal R} \over \mu} \right)^{4-\delta}
\Omega^{2\delta-5} \right)^{1 \over 6+\gamma-2\delta}.
\label{es}
\ende
Integrating  Eq. (\ref{phi}) over $z$ one finds
\beg
\int_0^1 \rho u_{R} dz = 2B^{\rm vert}B^{\rm tor},
\label{cond}
\ende
so that (\ref{Mdot}) turns into 
\beg
\dot M=-8\pi R \tilde{\rho} \tilde{c}_{\rm s} H B^{\rm vert} B^{\rm tor}.
\ende
The accretion is thus provided by the surface toroidal magnetic
field in our model in strong contrast to the standard-alpha disc theory. 
According to the latter, the surface density $\Sigma_0$ of
the disc is   
$\Sigma_0 = \dot M / 3 \pi \nu$
with $\nu= 2\alpha_{\rm SS} c_{\rm s} H_0$ and $c_{\rm s} = 2H_0 
\Omega_{\rm Kep}$,
where $\dot M$ and $H_0$ are the accretion rate and the disc half-thickness
without a magnetic field.\footnote{We shall  always neglect the 
factor $1-(R_*/R)^{1/2}$ with $R_*$ as the central object radius.}
The $\Sigma_0$ and $H_0$ as functions $M_*$, $R$ and $\dot M$ are
\begin{eqnarray}
\lefteqn{H_0=}\nonumber\\
\lefteqn{ {1 \over 2} \left( { 9 \over 32\pi (3\pi \alpha_{\rm SS})
^{1+\gamma}} 
{k_0 \over \sigma} 
\dot M^{2+\gamma} \Omega_{\rm Kep}^{2\delta-7-\gamma} 
 \left( {{\cal R} \over \mu} \right)^{4-\delta}
\right)^{1 \over 10-2\delta+3\gamma}, }
\end{eqnarray}
\begin{eqnarray}
\lefteqn{\Sigma_0={1 \over 3\pi} \left( \left( { 32\pi (3\pi)^{1+\gamma} 
\over 9}
{\sigma \over k_0} \right)^2 
\alpha_{\rm SS}^{2\delta-8-\gamma} \dot M^{6-2\delta+\gamma}
\Omega_{\rm Kep}^{4-2\delta-\gamma}\right.}\nonumber\\  
&&\left.\left( {\mu \over {\cal R}} \right)^{8-2\delta}
 \right)^{1 \over 10-2\delta+3\gamma}.
\end{eqnarray}
We are here interested to know the magnetic influence on the accretion
rate, the disc height and the surface density.
We shall consider two ratios for each quantities,
for
given accretion rate or for given surface density, resp..
Nevertheless, all these characteristics are related to each other, 
and consideration of one ratio is sufficient. For given accretion rate, it 
results
\beg
{H \over H_0}=2 \left( {2\pi (3\pi \alpha_{\rm SS})^{1+\gamma} \over 3}
\left( {\epsilon \over 4\pi \hat {\dot M}} \right)^{\gamma+2}
\right)^{1 \over 10-2\delta+3\gamma},
\ende
\begin{eqnarray}
\lefteqn{{\Sigma \over \Sigma_0}=}\nonumber\\
\lefteqn{2 \hat \Sigma \left( { 9 \over 4\pi^2
(3\pi )^{2\gamma+2} } \alpha_{\rm SS}^{8-2\delta+\gamma}
\left( {\epsilon \over 4\pi \hat {\dot M}} \right)^{6-2\delta+\gamma}
\right)^{1 \over 10-2\delta+3\gamma}}
\end{eqnarray}
and for given surface density  
\beg
\Psi={\dot M \over \dot M_0}= {4 \over 3 \epsilon } \hat{\dot M} 
\left( {4 \over 81} \alpha_{\rm SS}^{2\delta-8-\gamma} \left(
2 \hat \Sigma \right)^{2\delta-10-3\gamma} 
\right)^{1 \over 6-2\delta+\gamma}.
\label{psi}
\ende
Our calculations will reveal the latter quantity always exceeding unity so that for 
given column density $\Sigma$ the related accretion rate $\dot M$ is higher for magnetised 
disc; and for given accretion rate the related column density is smaller than 
without magnetic field.

The Mach number of the radial flow is
\beg
{\rm Ma}={u_{R} \over c_{s}}=\left({3 \over 5}\right)^{1/2}
 {u_{R} \over T^{1/2}},
\ende
where we have used the ideal gas sound speed $c_{\rm s}^2=5/3\ P/\rho$.

The inclination angle $i$ of the radial magnetic field component at the surface 
$B_{R{\rm s}}$ to the rotation axis
is 
\beg
\tan i = {B_{R\rm s} \over B^{\rm vert}}.
\label{tani}
\ende
Its determination for prescribed accretion rate is the central point of the 
present paper, the often formulated dragging problem corresponds to the theory of the vertical 
structure of accretion discs in external magnetic fields.

Below we shall use the constants $R=10^{10}$cm,
$ M_*= 1 M_{\sun} $, $\mu =0.6$, $k_0=6.6 \cdot 10^{22}$, $\gamma=1$,
$\delta=-7/2$.
The Kepler velocity is $\Omega_{\rm Kep}=1.2 \cdot 10^{-2}$s$^{-1}$. 
The normalisation constants
with $\epsilon=0.01$ as a typical value\footnote{It can be easy calculated
according to (\ref{em}) or (\ref{es}) and models values specified above
and in Table 1.} are
$\tilde \rho = 2.8 \cdot 10^{-8}$ g/cm$^{3}$,
$\tilde c_{\rm s} = 1.1 \cdot 10^6$ cm/s,
$\tilde T=8.7 \cdot 10^3$ K, $\tilde \nu =\tilde \eta= 1.1 \cdot 10^{14}$
cm$^2$/s, $\tilde F =3.7 \cdot 10^{10}$ erg cm$^{-2}$s$^{-1}$
and the unit for the magnetic field is $\tilde B=590$ G.

\begin{figure}
\psfig{figure=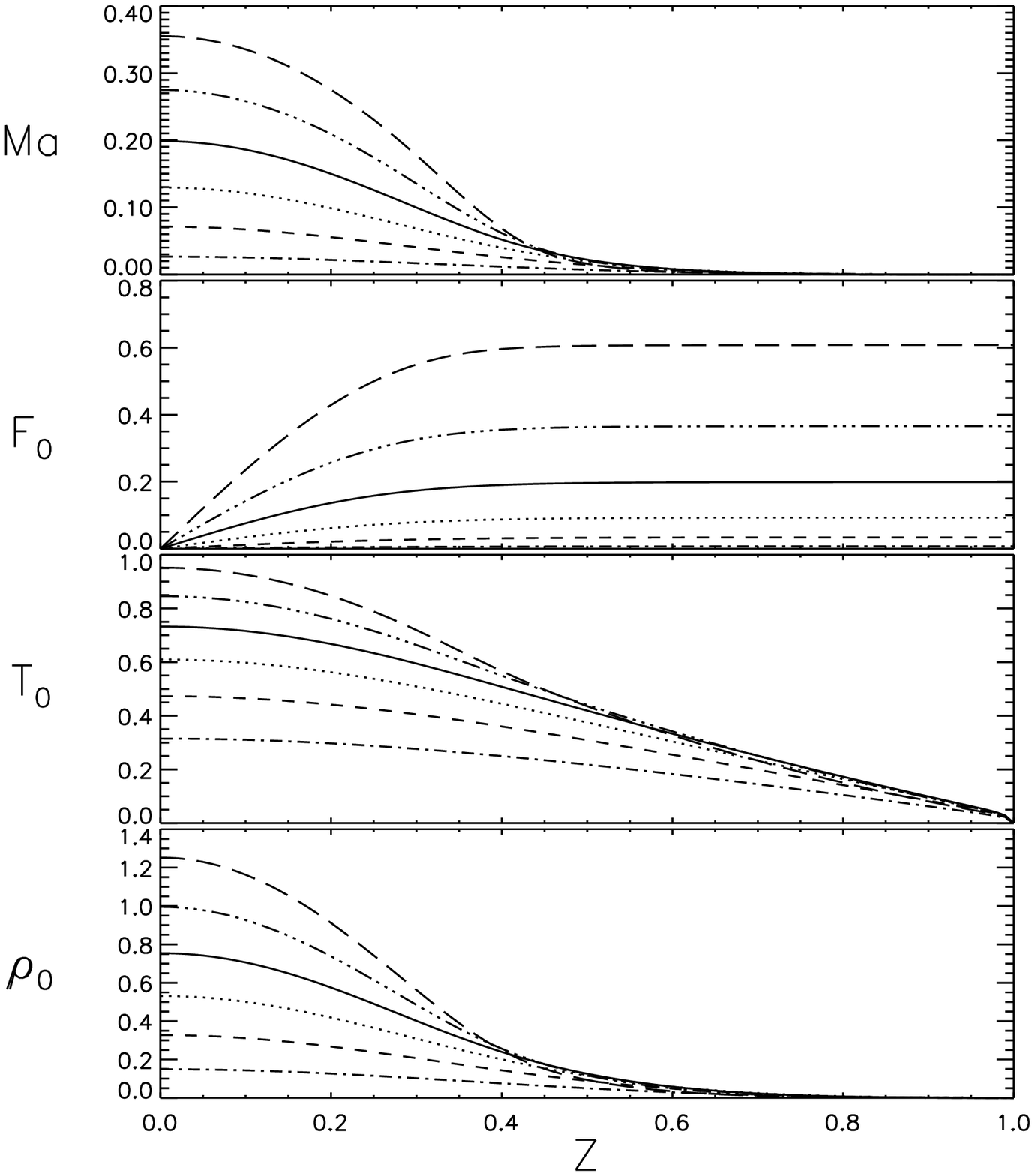,width=9cm,height=7cm}
\psfig{figure=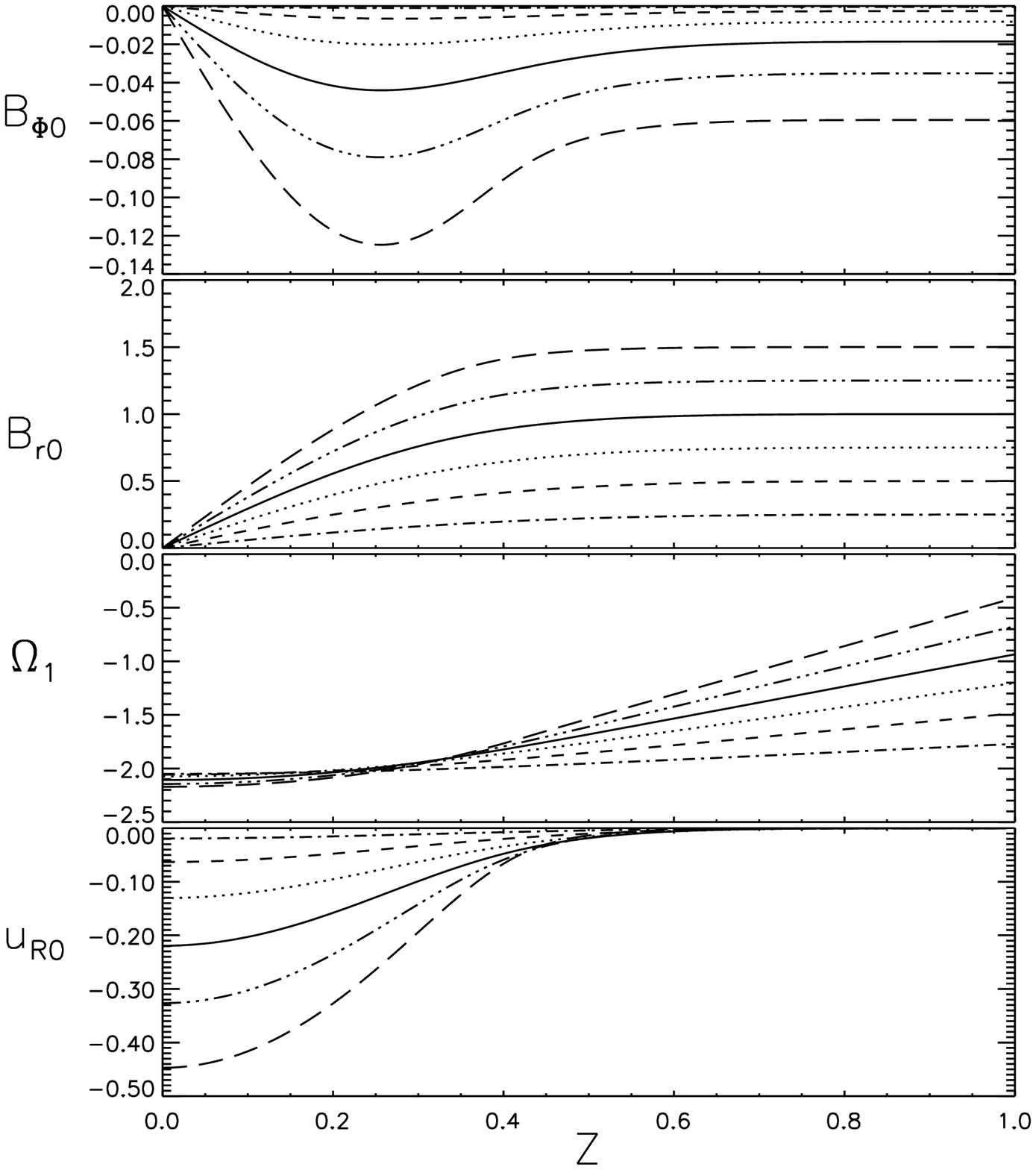,width=9cm,height=7cm}
\caption{The vertical disc structure for $\alpha_{\rm SS}=0.1$,
$ \Pm=1$ and $B^{\rm vert}=1$.
The $\beta$ is $-4 \cdot 10^{-4}$ (dot-dashed),
$-2.6 \cdot 10^{-3}$ (short-dashed),
$-8.2 \cdot 10^{-3}$ (dotted), $-1.9 \cdot 10^{-2}$ (solid),
$-3.5 \cdot 10^{-2}$ (dot-dot-dot-dashed),
$-5.1 \cdot 10^{-2}$ (long-dashed). The values of
$ \hat {\dot M}$, $\hat \Sigma$ and $i$ for each curve are
given in Table \ref{mdsig}}
\label{fig1}
\end{figure}
\begin{figure}
\psfig{figure=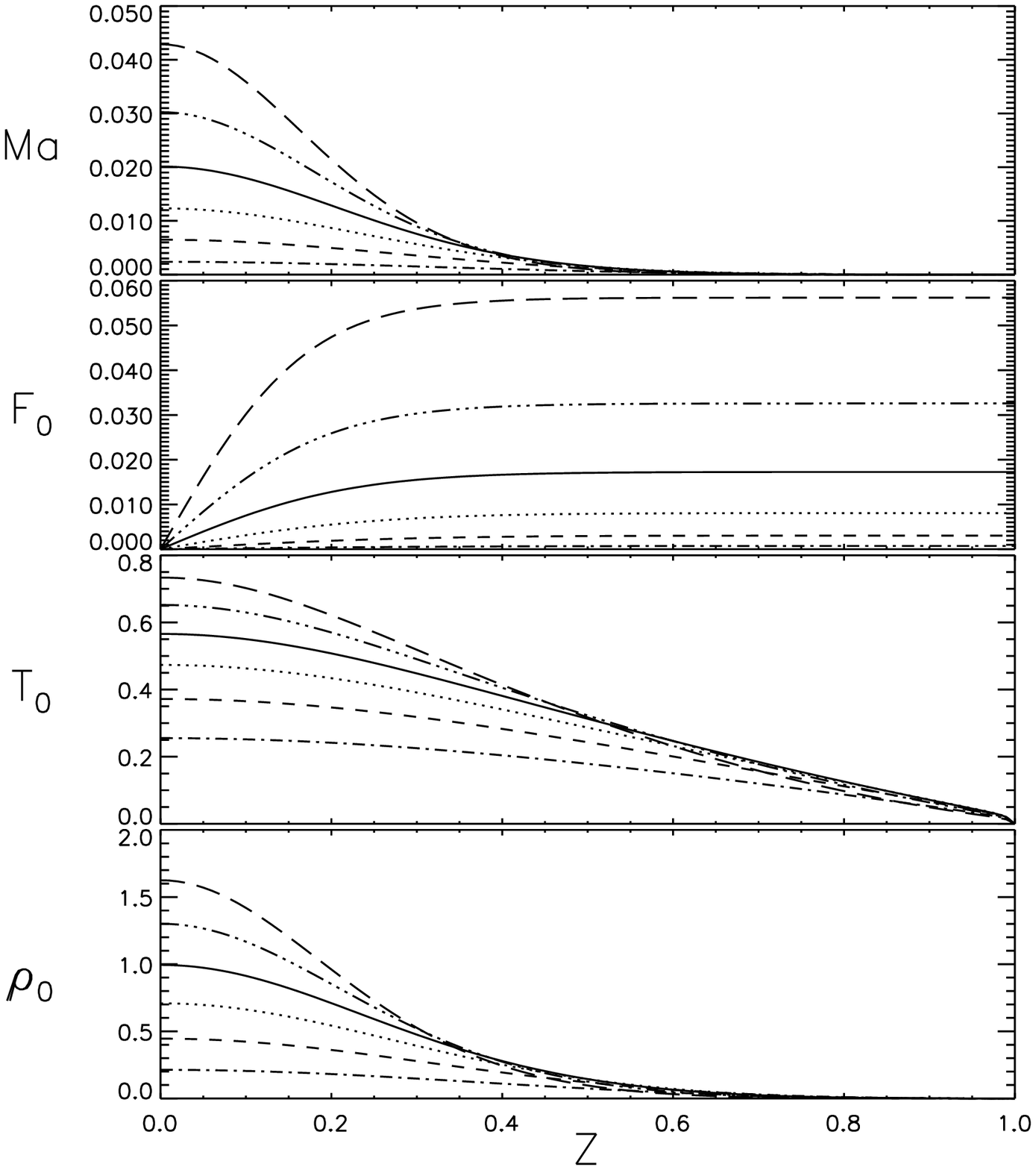,width=9cm,height=7cm}
\psfig{figure=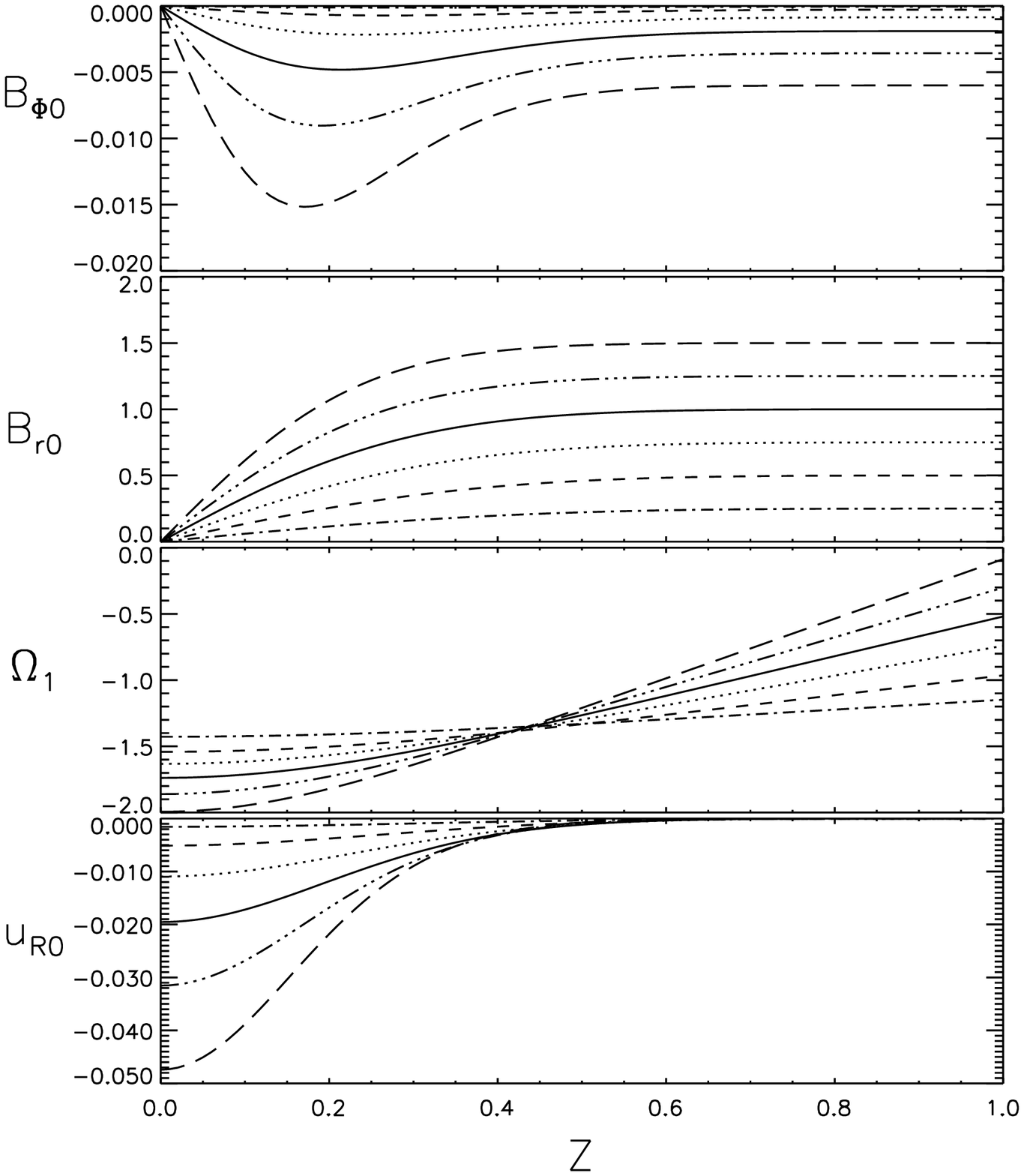,width=9cm,height=7cm}
\caption{The same as on Fig.\ref{fig1}, but for \ass$=$0.01} 
\label{fig2}
\end{figure}

\section{Results and discussion}
The systems of ordinary differential equations (\ref{uR0})$\dots$(\ref{F0})
with boundary conditions (\ref{b0}) and (\ref{b1}) are solved by a
relaxation method (see Press et al. 1992).
The dependent variables are $P$, $T$, $F$,
$u_{R}$, $u_{\phi}$, $B_{R}$, and $B_{\phi}$. The parameters are
$\alpha_{\rm SS}$, $\Pm$, $\dot M$, $B^{\rm vert}$ and $B^{\rm tor}$.
Only  Kramers opacity is used.

The calculation demonstrated that for a fixed $B^{\rm tor}$ one always can adjust $B_{R\rm s}$ to fulfil the condition (\ref{cond}). On the other hand, for fixed $B_{R{\rm s}}$ there is some critical value of
the $\alpha_{\rm SS}$  for given $\Pm$, $B^{\rm vert}$
and $B^{\rm tor}$ fixed by (\ref{cond}). It was possible to find the solution
only for $\alpha_{\rm SS} < \alpha_{\rm cr}$.  
The $\alpha_{\rm cr} \approx 0.4$ for $\Pm=1$, $B^{\rm vert}=1$ and 
$B_{R{\rm s}}=1$.
Due to the nonlinear character of the system the reasons for this numerical
problems are difficult to understand. Nevertheless, the real physical
interest requires calculations with $\alpha_{\rm SS} < 1$. 
We have restricted ourselves to calculations for $\alpha_{\rm SS}
\leq 0.1$ only. 

The vertical disc structure is illustrated in Figs. \ref{fig1} and
\ref{fig2}. We performed the calculations for vertical disc structure
as function of $\Pm$ and $B^{\rm vert}$ for one of three parameters
($\dot M$, $\Sigma$ and $i$) fixed. 
Two values of viscosity-alpha,
i.e. $\alpha_{\rm SS}=0.01$ and \ass =0.1, are used always. 
The qualitative behaviour of the solutions is very similar for  
both cases.
The profiles in Figs. \ref{fig1} and \ref{fig2} are labelled by
the toroidal inclination $\beta$. We could them also label  
either with the nondimensional accretion rate $\hat {\dot M}$ or the
nondimensional surface density $\hat\Sigma$ according to Table 1.

\begin{table}
\caption[]{The model parameters (Pm$=1$, $B^{\rm vert}=1$)} 
\begin{flushleft}
\begin{tabular}{|r|ccll|}
\hline
\ass & $\beta$ & $i$ & $\hat {\dot M}$     &    $\hat \Sigma$ \\ 
\hline\hline
0.1 & -4.0$\cdot 10^{-4}$ & 14$^\circ$ & 8.0$\cdot 10^{-4}$    & 0.062 \\
0.1 & -2.6$\cdot 10^{-3}$ &   27$^\circ$ & 5.2$\cdot 10^{-3}$  & 0.13  \\
0.1 & -8.2$\cdot 10^{-3}$ &   37$^\circ$ & 0.016               & 0.19  \\  
0.1 & -1.9$\cdot 10^{-2}$ & 45$^\circ$ & 0.037              &   0.25  \\
0.1 & -3.5$\cdot 10^{-2}$ & 51$^\circ$ & 0.070              &   0.31  \\
0.1 & -5.9$\cdot 10^{-2}$ & 56$^\circ$ & 0.12               &   0.37  \\
\hline
0.01 & -4.7$\cdot 10^{-5}$ & 14$^\circ$ & 9.3$\cdot 10^{-5}$ &   0.090 \\
0.01 & -2.8$\cdot 10^{-4}$ & 27$^\circ$ & 5.6$\cdot 10^{-4}$ &   0.17  \\
0.01 & -8.5$\cdot 10^{-4}$ & 37$^\circ$ & 1.7$\cdot 10^{-3}$ &   0.25  \\
0.01 & -1.9$\cdot 10^{-3}$ & 45$^\circ$ & 3.8$\cdot 10^{-3}$ &   0.31  \\
0.01 & -3.6$\cdot 10^{-3}$ &  51$^\circ$ & 7.1$\cdot 10^{-3}$ &  0.37  \\
0.01 & -6.0$\cdot 10^{-3}$ & 56$^\circ$ & 1.2$\cdot 10^{-2}$ &   0.41  \\
\hline
\end{tabular}
\end{flushleft}
\label{mdsig}
\end{table}

Almost the entire variation of the disc variables
happens for $0<z<0.5$ hence always we have `atmospheres' for $z>0.5$.
The atmosphere becomes thinner for increasing values of the vertical field.
The flow is subsonic ($\rm{Ma}<1$) and the applied surface
toroidal magnetic field is much smaller in comparison with $B^{\rm vert}$ 
($|\beta| \leq 0.1$) for all calculated cases. 
The density and the temperature weakly depend on the turbulence
parameters \ass\ and $\Pm$ for large \ass. 
The accretion flow at the disc midplane can be comparable
with the sound velocity. However,
due to the weak  dependence of the temperature on \ass, the Mach number behaves almost
linear with $\alpha_{\rm SS}$ and therefore $\rm{Ma} \ll 1$ for \ass$\ll 1$. 

\subsection{Accretion rate and surface density}

\begin{figure}
\psfig{figure=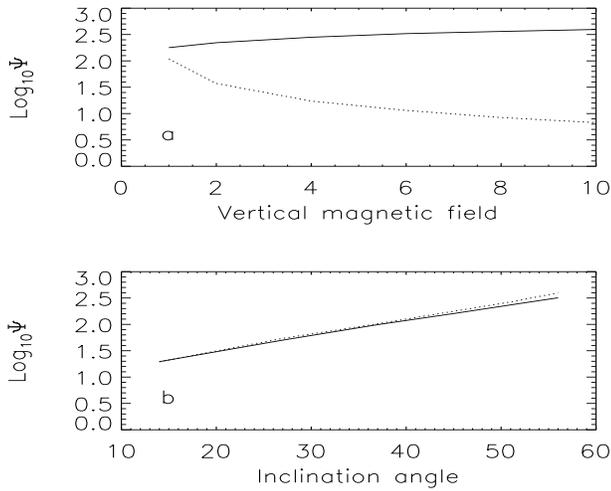,width=9cm,height=7cm}
\caption{The magnetic-disc accretion rates (normalized  with
 accretion rates without magnetic fields)  as a function of {\bf a)} the 
 vertical field $B^{\rm vert}$ for  \ass=0.01, 
$\beta= -2\cdot 10^{-3}$ (solid) and $ \Sigma=1.4$ g/cm$^2$ (dotted);
{\bf b)} the inclination angle  $i$ for
$B^{\rm vert}=1$, \ass=0.1 (solid), \ass=0.01 (dotted).
$\Psi$ is defined
by (\ref{psi}). Pm=1}
\label{fig4}
\end{figure}
\begin{figure}
\psfig{figure=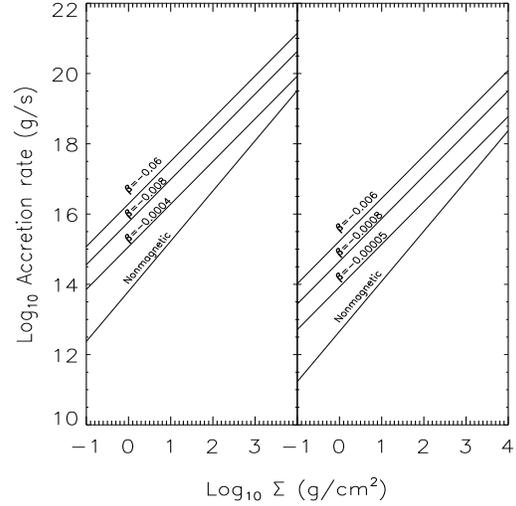,width=8cm,height=7cm}
\caption{The accretion rate  as a function of the surface density  $\Sigma$ for
$\alpha_{\rm SS}=0.1$ (left) and  $\alpha_{\rm SS}=0.01$ (right).
The numbers are given in Table 1. Pm$=$1, $B^{\rm vert} = 1$}
\label{fig5}
\end{figure}

The accretion rate for magnetic discs  drastically increases
in comparison with nonmagnetic discs (see Figs.
\ref{fig4} and \ref{fig5}).
Nevertheless, the accretion rate for magnetic
discs can be even smaller than for nonmagnetic discs for large
$\Pm$ and small $i$ (the last case is the same as the case
with large $B^{\rm vert}$ when $\hat \Sigma$ fixed).

Due to the direct relation between accretion rate and surface density
(Fig. \ref{fig5}) the dimensionless surface density dependence
on parameters is the same as for the dimensionless accretion rate.
The surface density for magnetic discs is usually smaller than for
nonmagnetic discs for the same accretion rate.
The magnetic disc radial velocity is
greater than nonmagnetic disc radial velocity (the accretion rate
is greater in magnetic disc). The surface density 
does not depend on $\alpha_{\rm SS}$ because the toroidal magnetic field
is always small and we can neglect its influence on the pressure. 

\subsection{Disc thickness}

\begin{figure}
\psfig{figure=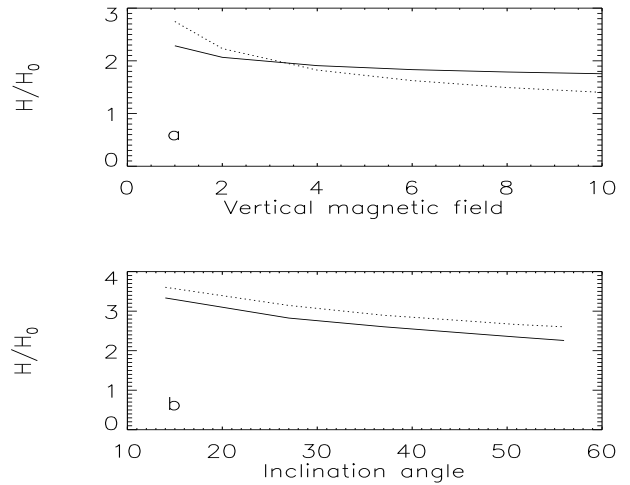,width=9cm,height=7cm}
\caption{The magnetic-disc height in comparison with
nonmagnetic-disc height as a function of
{\bf a)} the vertical magnetic field  $B^{\rm vert}$. 
$\beta= 2\cdot 10^{-3}$ (solid) and $\dot M=4\cdot 10^{16}$ g/s (dotted), 
\ass=0.01; {\bf b)}
the inclination angle  $i$ for
 \ass=0.1 (solid), \ass=0.01 (dotted).
Pm=1, $B^{\rm vert}=1$}
\label{fig6}
\end{figure}

In Fig. \ref{fig6} the thickness 
for magnetic discs is shown in comparison to the nonmagnetic case. The magnetic disc is 
    thicker
than the nonmagnetic one for all calculated models. Nevertheless, this
fact is only a rule with exceptions. 
Note that magnetic discs can be thinner than nonmagnetic discs for
both large $B^{\rm vert}$ and $\Pm$. However, we should
 care the radial flow Mach number. The radial flow can become
supersonic for $H < H_0$ for large enough
$\alpha_{\rm SS}$. This problem completely disappears for small $\alpha_{\rm SS}$.
The disc thickness decreases with increasing radial magnetic field and vertical
magnetic field due to magnetic stresses (see Fig. \ref{fig6}). It is
interesting to note that the accretion rate is getting larger for thinner disc.
The disc thickness weakly depends on Pm as well as \ass.

\subsection{Inclination angle $i$}

\begin{figure}
\psfig{figure=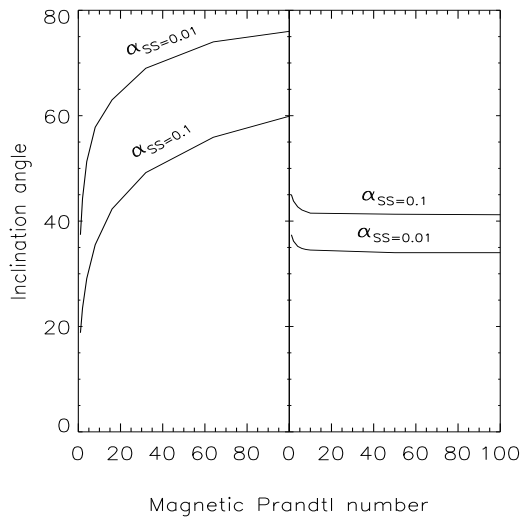,width=8cm,height=7cm}
\caption{The inclination angle $i$ of magnetic field lines to the rotation 
axis 
 for $B^{\rm vert}$=1. {\bf LEFT}: toroidal field fixed  ($\beta=-9 \cdot 
    10^{-4}$), {\bf  RIGHT}: column density fixed 
$\Sigma$ =1.4 g/cm$^2$}
\label{fig7}
\end{figure}

We find that  for rather low $\beta$ the inclination angle $i$ can be larger 
than the critical value of $30^\circ$ (Blandford \& Payne 1982) for jet 
launching even for Pm of order unity. This result is in accordance with our
calculation for polytropic magnetic disc (paper I). 
Moreover, fog given surface density the inclination angle hardly depends
on Pm (Fig.\ref{fig7}) also in accordance with the results of paper I.
This means that the radial velocity is proportionate
to 1/Pm in this case. Nevertheless, the inclination angle depends
strong on Pm for given accretion rate. 
The inclination angle increases for decreasing $\alpha_{\rm SS}$ for fixed
accretion rate and decreases for fixed surface density. Such a
behaviour is the result of the behaviours of accretion rate and
surface density with changing $\alpha_{\rm SS}$.

According to (\ref{cond}) and (\ref{Mdot}), the accretion rate is connected
with $\beta$. So, for given $B^{\rm vert}$ 
there is a direct relation between the  $\beta$  
and the resulting inclination angle $i$, Table 1 present the numbers. 
Our model yields high accretion rates and high inclination angels already for 
rather small $\beta$. The larger the $\beta$  is the higher the $i$. 
According to (\ref{camp}), the $\beta$ increased inwards.
We can  conclude that the Blandford-Payne condition is fulfilled most 
easily for the inner part of the accretion disc. 

\section{Conclusion}
The vertical structure of accretion discs with an imposed vertical magnetic
field has been considered.
The magnetic field energy is  supposed as to be in balance with the 
accretion flow energy. The angular momentum transport  is  fully provided
by the magnetic field, the Reynolds stress does here not play any role.

The turbulent viscosity
might  only be important for the energy balance of  the disc but the 
calculations demonstrated also the dominance of the Joule heating. The
magnetic field is, therefore, the essential feature of the  model and
the equations do not change to standard $\alpha$-disc equations in the
small magnetic field limit. Stehle \& Spruit (2001) results is also
confirmed the existence of magnetic field induced accretion.
The  results  actually demonstrate the close relation between
magnetic field dragging and the vertical structure of thin
accretion disc. 
The interaction of the magnetic field with the disc can drastically 
change the disc structure as well as the configuration of the
magnetic field. The angular velocity differs  from the Keplerian one. 
This difference  is, however, relatively small because the magnetic 
energy is small compared with  the gravitational energy in our model.
The radial velocity, however,  can increase drastically 
becoming  comparable to the sound speed for some models.
This can lead to a strong  amplification 
of the accretion rate for a given column density (Fig.\ref{fig5}).

The accretion rate is connected directly with the surface toroidal
magnetic field in our model. Nevertheless, even small  
toroidal magnetic field at the surface ($\beta \ll 1$) is enough 
to increase the accretion rate drastically. The accretion rate can
be as amplified drastically (large $i$ and small Pm), so suppressed
(small $i$ and large Pm).

The magnetic disc is basically thicker than the nonmagnetic one for 
typical parameters values. Nevertheless, we can not exclude that the magnetic
disc can became thinner than a nonmagnetic one for some parameters 
(e.g. large $B^{\rm vert}$).

However, the most surprising results of our calculations  concern the
inclination angle $i$ of the magnetic field lines to  the 
rotation axis.  We found that i) already  rather small toroidal field
component $B^{\rm tor}$ can produce inclinations exceeding 
the critical value of $30^\circ$  and ii) this
effect is almost independent of the magnetic Prandtl number 
for given surface density Fig. ({\ref{fig6}).

In previous studies (Lubow, Papaloizou \& Pringle 1994; Reyes-Ruiz
\& Stepinski 1996)  large radial inclinations  could only  be obtained  for
${\rm Pm} \ge 100$. After our results it also holds for Pm=1. The 
difference is due to the fact
that the previous studies neglect the magnetic field influence on disc
structure and used radial velocity from standard accretion disc
theory.

For a given accretion rate, $\dot M$,  the larger  $\beta$  lead to 
the larger   $i$ and to the smaller $\Sigma$ (Fig. \ref{fig5}). 
As after (\ref{camp}) the larger (negative)  $\beta$ exist in the
innermost accretion disc region, we have there the smaller column 
density and the  stronger  radial inclination of the field lines. 
The jet launching should thus be concentrated to the inner region
of an accretion disc.   

Only Kramers opacity was used in this paper but it might be very
interesting to apply other opacity laws to reformulate the stability
problem of the structure of accretion discs.

\bl
\noindent {\bf Acknowledgements:}
D.S. thanks for the kind  financial support by the Deutsche 
Forschungsgemeinschaft and Russian foundation for basic research
(grant RFBR-DFG 00-02-04011). G.R.  acknowledges the kind 
hospitality of the HAO  during the work on this paper.
 
{}

\end{document}